**Interfacial origin of magnetic compensation and topological-like Hall effect in crystalline lutetium iron garnet thin films**

*Satakshi Pandey*[1,*], *Anne Forget*[1], *Sarah Clement*[2], *Christophe Gatel*[2], *Bénédicte Warot-Fonrose*[2], *Grégoire de Loubens*[1], *Jean-Baptiste Moussy*[1,*]

[1] Université Paris-Saclay, CEA/CNRS, Service de Physique de l'Etat Condensé, 91191 Gif-sur-Yvette, France.

[2] Université de Toulouse, CNRS, CEMES, 31055 Toulouse, France.

E-mail: (satakshi.pandey@cea.fr, jean-baptiste.moussy@cea.fr)

## Abstract

Ferrimagnetic oxides such as rare-earth iron garnets (REIG) have emerged as essential components of spin-based devices owing to their high Curie temperatures, low Gilbert damping, and excellent insulating properties. Recently, REIGs have attracted immense interest for hosting robust chiral spin textures that can be potential information carriers in next-generation spintronic devices. Here we investigate the structural, chemical, and magnetic behaviour of single-crystal $Lu_3Fe_5O_{12}$ (LuIG) thin films for different film thicknesses on a $Gd_3(Mg,Zr)Ga_5O_{12}$(111) or SGGG(111) substrate. Our results demonstrate the presence of a 3 nm-thick Gd-rich LuIG layer at the LuIG/SGGG interface, which gives rise to a magnetic compensation temperature around 50 K. We further probe the anomalous Hall effect (AHE) in LuIG layers with an adjacent platinum ultrathin film. In addition to the magnetic compensation point, the AHE loops show pronounced nonmonotonic features, often attributed to the spin-Hall topological-like Hall effect (SH-THE) and non-collinear spin textures arising from the Dzyaloshinskii-Moriya interaction (DMI). LuIG thickness-dependent AHE measurements reveal that the composition variation due to Gd diffusion breaks the inversion symmetry and could contribute to DMI. These findings shed light on fundamental understanding and engineering of the different interfaces of magnetic garnets for advanced spintronic device applications.

## 1 Introduction

Ferrimagnetic insulators such as rare-earth iron garnets (REIGs) with the general formula $RE_3Fe_5O_{12}$ have attracted a considerable amount of interest, particularly due to fast current-driven dynamics, no Ohmic loss, and large magnon propagation length, making them ideal for signal transmission and magnon logic devices [1–3]. REIG films exhibit ultralow damping, high Curie temperature ($T_c$ > 500 K), and tunable perpendicular magnetic anisotropy (PMA), facilitating their integration into magnetic memory and sensor technologies [4–6]. In addition, REIG films can show several interesting spintronic phenomena, including the proximity effect [7][, spin pumping[8], and the spin Seebeck effect [9–11] when incorporated into heterostructures with a heavy metal (HM), for example, platinum [2] or tungsten [12]. HM/REIG hybrid systems are used in spin-orbit torque (SOT)-induced magnetisation switching at low current density, paving the way for magnetic insulator-based ultralow-dissipation SOT devices [13–17].

Among various REIG systems, yttrium iron garnet (YIG) and lutetium iron garnet (LuIG) possess the lowest magnetic damping, S-state magnetic ions, small anisotropy, and the highest magnetic saturation, making them highly appealing for efficient spin-wave transport[18–20]. Despite their desirable properties, the growth of high-quality epitaxial YIG or LuIG thin films is typically restricted to epitaxial garnet substrates such as gadolinium gallium garnet ($Gd_3Ga_5O_{12}$ or GGG) or substituted-GGG ($Gd_3(Mg,Zr)Ga_5O_{12}$ or SGGG). The growth process requires very high temperatures or post-growth annealing that could reach more than 700 $^{\circ}$C, which can lead to cation intermixing at the interface [21–23]. This, in turn, could modify magnetic properties at the interface due to incorporation of paramagnetic Gd ions into the diamagnetic rare-earth sublattice sites [24]. The interaction between Fe and Gd ions in the interfacial layer can also give rise to a compensation temperature because of the Gd *4f* shell magnetic moment, which becomes dominant at low temperatures [21]. Furthermore, the compensation temperature can be modulated by varying the Gd content in the REIG film [25]. Interestingly, recent reports have shown that the inhomogeneous Gd concentration across the thickness of magnetic films can break the structural inversion symmetry, promoting Dzyaloshinskii–Moriya interaction (DMI), a key ingredient for stabilizing nontrivial spin textures [26–28].

In the present work, we investigated the crystalline growth of LuIG thin films of varying thicknesses ($t$) (3 nm < $t$ < 20 nm) using pulsed laser deposition (PLD) on SGGG(111) substrates. We performed a comprehensive study of the structural, chemical, and magnetic properties of the LuIG thin films as a function of film thickness. Transmission electron microscopy (TEM) and electron energy loss spectroscopy (EELS) measurements revealed that a 3 nm-thick Gd-rich LuIG layer, $(Lu_{1-x}Gd_x)_3Fe_5O_{12}$, is formed at the film/substrate interface during the high-temperature growth process. Using magnetic and electronic transport measurements, we probed an unusual magnetic compensation point close to 50 K in LuIG thin films due to Gd-doping. Moreover, in the high-temperature regime, we detected additional peaks often associated with the spin-Hall topological-like effect (SH-THE), possibly arising due to DMI. The development of crystalline LuIG thin films with low damping, magnetic compensation, and SH-THE-like effect makes them appealing for spintronic applications.

## 2 Results and Discussions

### 2.1 Growth and Structural Analysis of Epitaxial LuIG Films

**Figure 1 (a)** shows a schematic description of the LuIG garnet, which exhibits a cubic structure in the bulk. Its bulk lattice constant is $a_{LuIG}$ = 1.228 nm. LuIG is a ferrimagnet with three magnetic sublattices, where the two Fe sublattices are strongly coupled via antiferromagnetic superexchange interactions. Each unit cell consists of 12 dodecahedrally ($c$) coordinated Lu atoms, 8 octahedrally ($a$) coordinated Fe atoms, and 12 trivalent Fe atoms that are tetrahedrally ($d$) coordinated with O atoms.

Epitaxial LuIG(111) films with varying thickness are grown on SGGG(111) substrate by pulsed laser deposition (PLD) as described in the Experimental section. Reflection high-energy electron diffraction (RHEED) patterns of the LuIG thin films are recorded along the $[11\bar{2}]$ azimuth before growth and during cool-down under molecular oxygen (see **Figure S1 (a)** and **(b)**). The RHEED patterns exhibit sharp streaks without any additional spots, revealing a two-dimensional growth mode of LuIG without secondary phases.

The surface morphology of the SGGG substrate and LuIG films is also examined via atomic force microscopy (AFM) (see **Figure S1 (c-e**), which confirms atomically flat surfaces with a low root-mean-square roughness (RMS). Additionally, we observed a few particles, also reported previously in the literature on garnet thin films [2]. The film thickness is obtained using X-ray reflectivity (XRR) measurements. **Figure S1 (f)** shows XRR measurements with adjustments of LuIG thin films for two different thicknesses deposited on the SGGG substrate. The roughness and thickness of the films are analyzed using Parratt's formalism. The fits yield to a RMS roughness below 0.5 nm for both 20 nm and 7 nm-thick thin films. At this stage, it is relevant to highlight that we have added an additional layer at the film/substrate interface, in order to obtain a good fit of the Kiessig oscillations. The additional layer has a lower density than that of LuIG, and it could be due to diffusion at the LuIG/SGGG interface during the high-temperature growth process [22]. The thickness of this interfacial region is around 3 to 4 nm for both samples. As we will describe below, additional structural characterisations (TEM and EELS) have been performed to assess the presence of this diffusion interface layer.

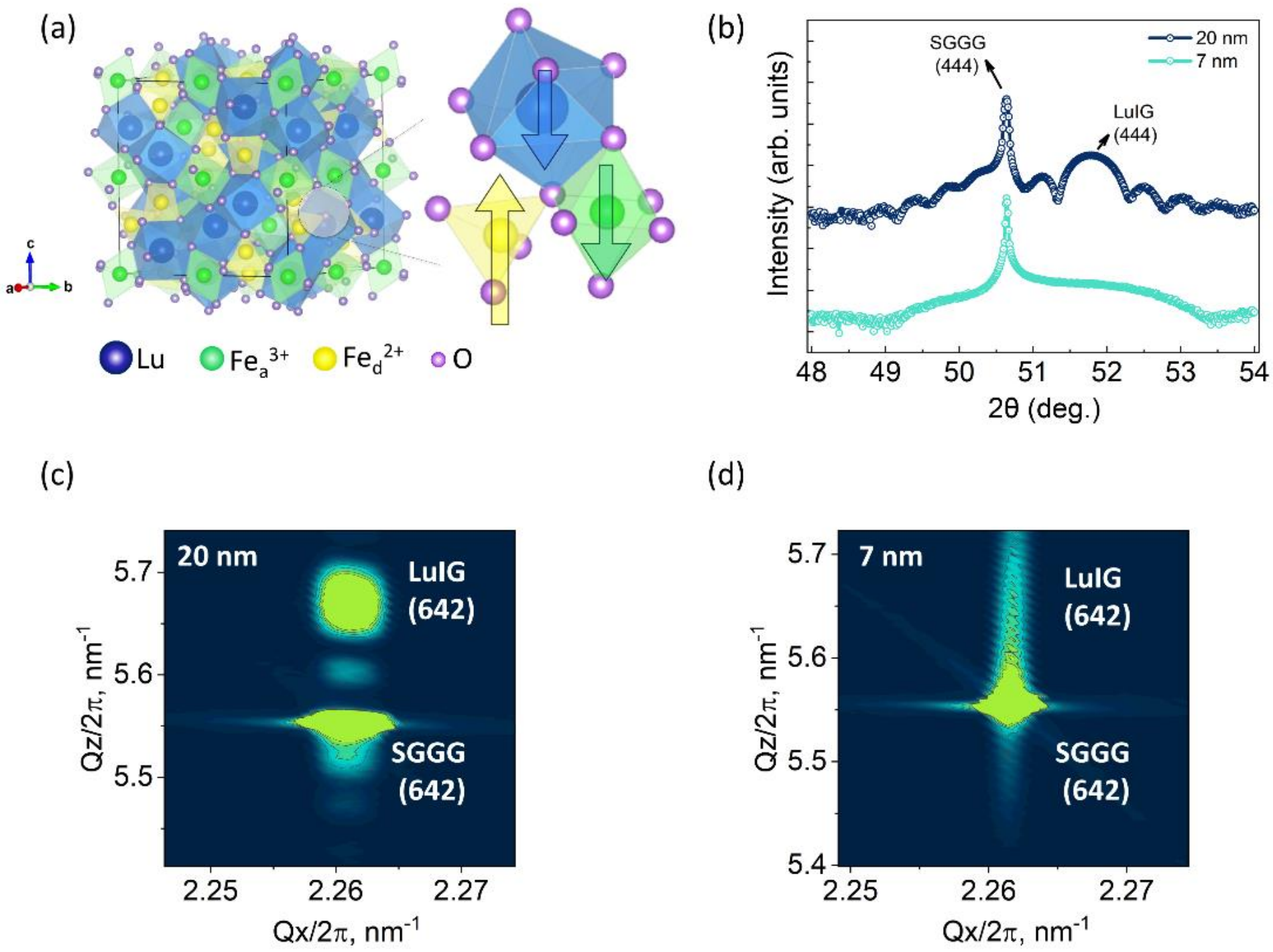

***Figure 1:*** *(a) Schematic of the lutetium iron garnet ($Lu_3Fe_5O_{12}$) crystal structure with the three different magnetic sublattices, where the $Lu^{3+}$ ion is dodecahedrally (blue)coordinated, while the $Fe^{3+}$ ions are tetrahedrally (yellow)and octahedrally (green)coordinated. (b) XRD $2\theta$–$\omega$ patterns of the 20 nm (blue) and 7 nm (green) films, showing the (444) Bragg peaks with clear Laue oscillations. (c,d) RSM patterns around the (642) asymmetric diffraction peak for the 20 nm and 7 nm LuIG layers*

X-ray diffraction (XRD) measurements shown in **Figure 1 (b)** provide additional information on the film structure. XRD $2\theta$-$\omega$ scans for the 20 nm and 7 nm films show only (*hhh*)-type reflections corresponding to the (111) crystallographic orientation, and clear Laue fringes revealing the excellent film quality. The lattice constant of the SGGG substrate ($a_s$) ~ 1.248 nm, which is larger than that of bulk LuIG ($a_f$) ~ 1.228 nm, induces a theoretical in-plane tensile lattice mismatch of 1.62 %, calculated using the standard formula: $\frac{a_s - a_f}{a_f}$ . The obtained lattice parameters from XRD measurements for 20 nm and 7 nm LuIG films are given in **Table 1**. For our films, we observe that the out-of-plane lattice parameter is lower than in the bulk LuIG sample.

**Table 1:** Structural parameters extracted from XRD measurements and reciprocal space maps around the (642) diffraction peak.

| Sample | SGGG substrate | | LuIG film | | | |
|---|---|---|---|---|---|---|
| | a (nm) | c (nm) | a (nm) | c (nm) | In-plane strain $\varepsilon_{\parallel}$ | OOP strain $\varepsilon_{\perp}$ |
| **LuIG (~ 7 nm)** | 1.2506 | 1.2475 | 1.2504 | 1.2163 | +1.82 % | -0.95 % |
| **LuIG (~ 20 nm)** | 1.2511 | 1.2478 | 1.2509 | 1.2216 | +1.86 % | -0.52 % |

We extract additional information on out-of-plane as well as in-plane lattice parameters from reciprocal space maps (RSM) along the (642) oblique plane. The RSM scans on the 20 nm thick film and 7 nm film are shown in **Figure 1(c)** and **1(d),** respectively. The measured lattice constant, IP, and OOP strain for the LuIG films are listed in **Table 1**. The lattice parameters obtained from RSM are consistent with the lattice constants measured by XRD $2\theta$–$\omega$ scans.

The OOP lattice parameters (c) deduced from RSM maps are also coherent with the lattice constants measured by XRD $2\theta$–$\omega$ scans. Therefore, the IP lattice parameter of the LuIG film is very close to that of the substrate because of strained growth on the SGGG substrate [Guo PRM 2022] while the OOP parameter is smaller than in bulk LuIG, indicating a compressive strain along the [111] direction. The OOP strains ($\varepsilon_{\perp}$) and IP strains ($\varepsilon_{\parallel}$) can be calculated by the following equations $\varepsilon_{\perp} = \frac{c-a_0}{a_0}$ *and* $\varepsilon_{\parallel} = \frac{a_{\parallel}-a_0}{a_0}$, respectively, where $a_0$is the relaxed bulk lattice parameter of the LuIG film (see Table 1). Both films are under OOP compression ($\varepsilon_{\perp}<$ 0) and biaxial tensile strain ($\varepsilon_{\parallel} > 0$) on SGGG. Recent reports on LuIG thin films grown on SGGG substrates have demonstrated that tensile-strained films can introduce a lack of inversion symmetry, which could potentially enable DMI-tunable spin dynamics in LuIG thin films [18,26].

We have further performed an atomic-scale investigation of the LuIG/SGGG interface to identify strain and diffusion areas using high-resolution transmission electron microscopy (HRTEM) and electron energy-loss spectroscopy (EELS). TEM lamellae of the samples were prepared using the focused ion beam (FIB) milling technique. .

The high-quality crystalline structure of a 20 nm thick LuIG film on the SGGG substrate, as well as the epitaxial film growth, is clearly evident in the HRTEM image shown in **Figure 2 (a)**. The LuIG film appears coherently strained and adopts the substrate's in-plane lattice parameter, confirming pseudomorphic growth of the LuIG thin film. Moreover, we do not observe any structural defects such as dislocations at the LuIG/SGGG interface.

To further investigate this behavior, we performed STEM-EELS line scans to determine the element distribution. Energy positions of the Fe L and Gd M edges have been mapped across the interface, as shown in Figure. **2(b) and 2(c)**, respectively. We clearly detect peaks of high intensity related to the presence of Fe and Gd atoms in LuIG and GGG, respectively. Moreover, we observe the coexistence of the Gd and Fe peaks in a single EELS spectrum in the LuIG film close to the interface. The Gd peak vanishes in the region far from the interface. These EELS analyses confirm the existence of a 3.5 nm thick Gd-rich layer at the LuIG/SGGG interface due to Gd diffusion from the substrate into the LuIG film. The formation of a Gd-doped layer was previously reported in the case of YIG thin films grown at high temperatures [17,22,27,28]. We also acquired HRTEM images of a 7 nm thin LuIG film capped with a 3 nm thick Pt layer (**Figure 2d**) deposited *in situ* by e-beam evaporation. The HRTEM image confirms the single-crystalline growth of the LuIG thin layer with a polycrystalline structure for the Pt overlayer, as expected for growth at room temperature. Two regions of the EELS spectrum from this sample is shown in **Figure 2(e) and 2(f)**. Again the clearly reveal the presence of a 3 nm thick Gd-rich LuIG layer at the film/substrate interface. Thus, TEM images and STEM-EELS analysis further validate the XRD and XRR results and reveal a strained structure in both film thicknesses, along with the formation of an interfacial Gd-doped layer. It is expected that the magnetic properties of the LuIG layer as a function of temperature would be influenced by the diffusion of Gd.

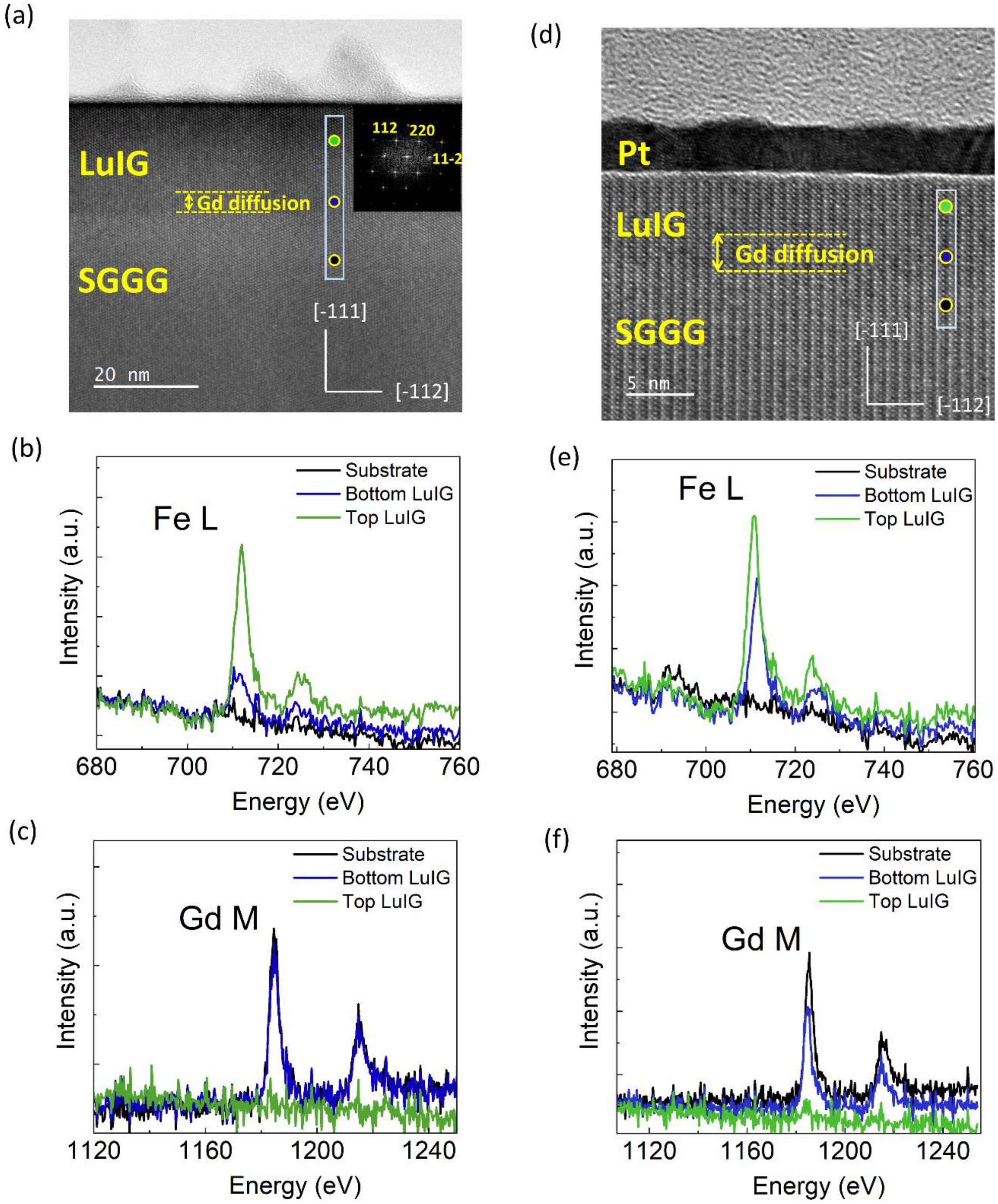

**Figure 2:** (a) High-resolution TEM (HRTEM) image of a 20 nm thick LuIG(111) film grown on a SGGG(111) substrate. The inset shows the corresponding fast Fourier transform (FFT) of the LuIG layer. (b and c) Fe L-edge and Gd M-edge regions from the corresponding EELS spectrum A clear interdiffusion of Gd is observed. (d) HRTEM image of a 7 nm-thick LuIG(111) film grown on SGGG(111) capped with a Pt (3 nm) layer. (e and f) Fe L-edge and Gd M-edge regions from the corresponding EELS spectrum. Again, a clear interdiffusion of Gd is observed.

### 2.2 Static and Dynamic Magnetic characterizations

To have a better understanding of the magnetic behaviour of the LuIG layers, we performed magnetic measurements using vibrating sample magnetometry (VSM). The strong paramagnetic background contribution from the SGGG substrate has been systematically subtracted. We first measured the IP and OOP magnetic hysteresis loops at room temperature of a 20 nm thick LuIG film as shown in **Figure 3 (a)**. The OOP curve is comparable to the IP curve, clearly indicating an increase in perpendicular magnetic anisotropy (PMA) of the LuIG film owing to the large in-plane strain $\varepsilon_{\parallel}$ (~ 1.8 %). The saturation magnetization ($M_s$) is approximatively 90 $kA{\cdot}m^{-1}$, which is smaller than the bulk value at room temperature (~ 140 $kA{\cdot}m^{-1}$), possibly due to the presence of the interfacial Gd-doped LuIG layer [28,29]. **Figure 3 (b)** shows the IP and OOP magnetic loops of a 7 nm-thick LuIG film at room temperature. We again observe that the OOP hysteresis loop is similar to the IP curve as in the case of the 20 nm thick film. However, its $M_s$ is lower (about 75 $kA{\cdot}m^{-1}$ at room temperature), which could probably be due to a higher contribution of the interfacial layer to total magnetisation in thinner films. We finally measured IP and OOP hysteresis loops of the 7 nm thin LuIG film capped with a 3 nm Pt layer (see **Figure 3 (c))** with a comparable IP and OOP behaviour and a saturation magnetisation similar to that of the bare 7 nm LuIG film.

The magnetic easy axis of the LuIG films spontaneously lies in-plane due to the shape anisotropy and the small magnetocrystalline anisotropy observed in LuIG films [29]. The presence of PMA in our epitaxial LuIG(111) thin films is therefore owing to the large in-plane strain $\varepsilon_{\parallel}$ (~ 1.8 %), which is stronger than the shape anisotropy.

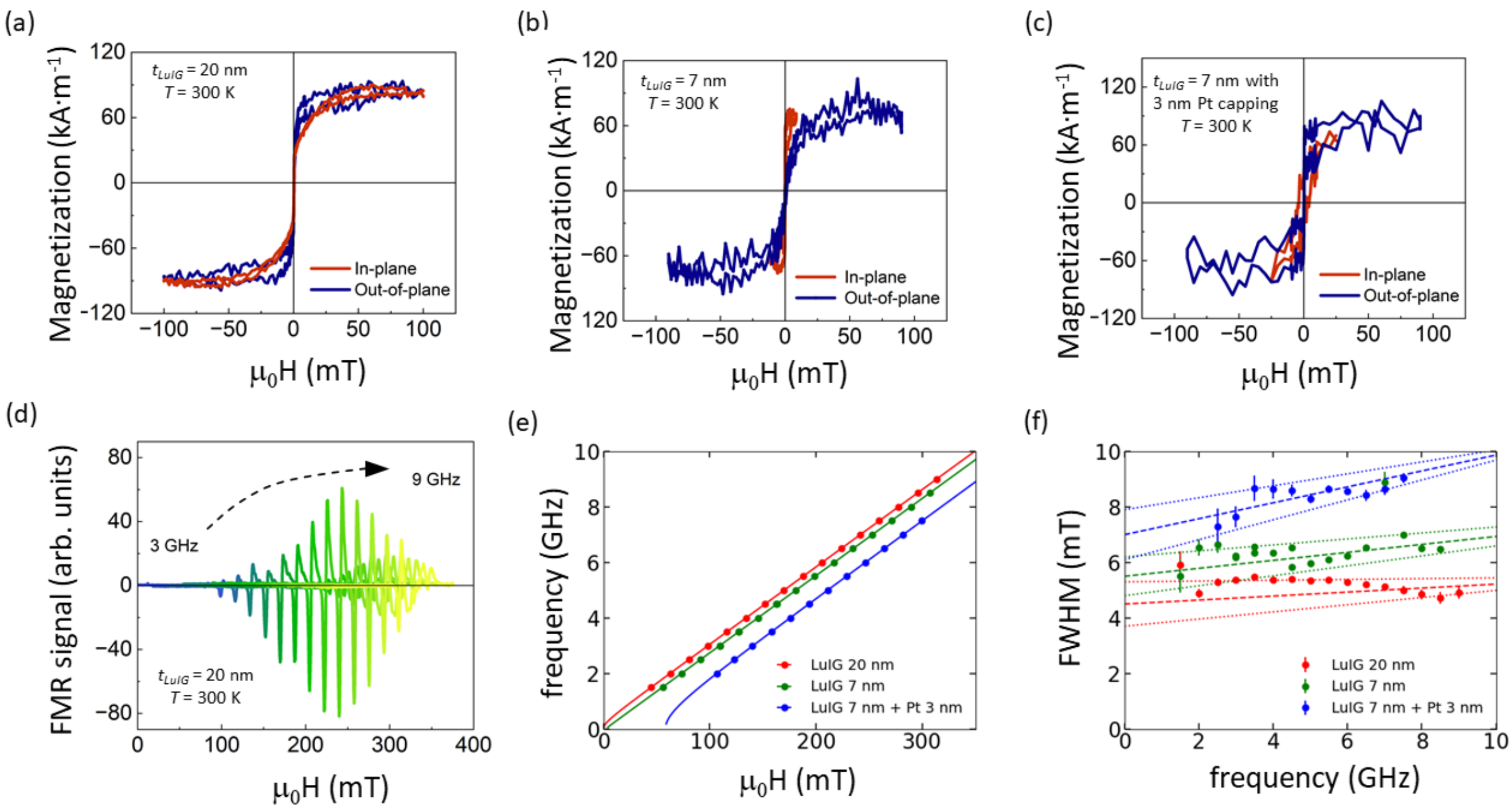

**Figure 3:** (a) In-plane (IP) and out-of-plane (OOP) hysteresis loops at room temperature of LuIG/SGGG samples with thicknesses of (a) 20 nm and (b) 7 nm and (c) 7 nm capped with Pt. (d) Room-temperature ferromagnetic resonance spectra of a 20 nm thick LuIG film at different frequencies from 3 GHz to 9 GHz. (e) FMR peaks plotted as a function of the applied magnetic

field for three different samples with the Kittel adjustments; (e) Linewidth dependence with frequency for the different films measured.

We next performed in-plane FMR measurements on the different LuIG films and Pt/LuIG bilayers using a broadband set-up operating at room temperature. The microwave power absorption spectra are recorded as a function of the applied bias magnetic field at various frequencies, as shown in **Figure 3 (d)** for the 20 nm thick LuIG film from 3 to 9 GHz. We fit them using a Lorentzian derivative function to determine the resonance field $H_{res}$ and the full linewidth at half maximum (FMHM) $\Delta H$ at each frequency. As often reported in the literature [30,31], our garnet films exhibit multiple resonance peaks that overlap, a signature that the ≈ 5 $mm^2$ area probed exhibits inhomogeneities of the internal magnetic field, most probably due to nonuniformities of PMA. Therefore, up to three peaks are used to adjust the FMR spectra (see **Figure S2 and S3**). **Figure 3 (e)** displays the dispersion relations of the main FMR peak for three different films, fitted using the Kittel equation, $f = \frac{\gamma\mu_0}{2\pi}\sqrt{H_{res}\left(H_{res}+M_{eff}\right)}$, where $\frac{\gamma}{2\pi} = \frac{g\mu_B}{\hbar}$ is the gyromagnetic ratio, $g$ is the Landé $g$-factor, $\mu_B$ is the Bohr magneton, $\hbar$ is the reduced Planck constant, and $M_{eff} = M_s - H_u$ the effective magnetization, equal to the difference between the saturation magnetization and the uniaxial perpendicular anisotropy field. The fitted values of the gyromagnetic ratio are independent of the film and correspond to $g$ = 1.991, very close to the free electron value as reported for LuIG [32]. On the contrary, the effective magnetization depends on the film. It is small and positive for the 20 nm thick film, corresponding to only 15% of the value of $M_s$ estimated from VSM. This means that a sizable $H_u$ reduces the shape anisotropy of this film; still, the latter does not exhibit net PMA. For the 7 nm film, it is slightly negative, i.e., $H_u$ is just large enough to compensate the shape anisotropy. For the 7 nm film covered with Pt, $M_{eff}$ has a larger negative value, i.e., the film has a net PMA ($H_u \approx 1.6M_s$). These findings are in agreement with the VSM cycles shown in **Figure 3 (a-c).** In FMR, the linewidth usually exhibits a linear dependence on frequency, which allows one to use the equation $\Delta H = \Delta H_0 + \frac{4\pi\alpha f_{res}}{\gamma\mu_0}$ to extract the Gilbert damping $\alpha$ and the inhomogeneous linewidth $\Delta H_0$. **Figure 3 (f)** shows the FWHM as a function of frequency for the three films. The linewidth of the 20 nm thick film, around 5 mT, is almost independent of frequency from 2 to 9 GHz. This means that it is completely dominated by inhomogeneities, and only an upper bound of the damping can be estimated for this film, $\alpha \lesssim 0.002$. Due to the presence of several nearby FMR peaks in the signal, the quality of the data sets for the 7 nm films is not excellent. Still, the FWHM of the bare 7 nm film is clearly about 1 mT broader than that of the 20 nm film and exhibits a slightly larger slope vs. frequency, corresponding to $\alpha \approx 0.002$. This slope is twice larger ($\alpha \approx 0.004$) for the 7 nm film covered with 3 nm of Pt, which can be ascribed to spin pumping due to the presence of Pt [33,34]. This latter film also exhibits the largest inhomogeneous contribution among the three films, a common observation for films exhibiting large PMA.

**Table 2:** Magnetic properties of LuIG(111) thin films on SGGG(111) substrates extracted from FMR and VSM analyses.

| Sample | $M_s$(kA·m$^{-1}$) | γ/(2π) (GHz/T) | $\mu_0 M_{eff}$ (mT) | α | $\mu_0 \Delta H0$ (mT) | $\mu_0 M_s$ (mT) | $\mu_0 H_u$ (mT) |
|---|---|---|---|---|---|---|---|
| **LuIG(20 nm)** | 90 ±5.5 | 27.87 ±0.01 | 19.4 ±0.2 | 0.001 ±0.0008 | 4.5 ±0.8 | 113 ±6 | 93.6 ±6.2 |
| **LuIG(7 nm)** | 75 ±7 | 27.84 ±0.03 | -3.2 ±0.4 | 0.002 ±0.0005 | 5.5 ±0.7 | 94 ±10 | 97.2 ±10.4 |
| **Pt(3nm)/LuIG(7 nm)** | 75 ±4 | 27.87 ±0.03 | -58.3 ±0.4 | 0.004 ±0.001 | 7.0 ±0.9 | 94 ±10 | 152.3 ±10.4 |

Overall, the magnetic parameters extracted from VSM and FMR measurements highlight the magnetic quality of our LuIG thin films and are summarized in **Table 2**.

### 2.3 Observation of compensation behaviour in LuIG thin films

We next measured the OOP hysteresis loops of the 20 nm-thick LuIG thin film as a function of temperature (**Figure 4 (a)**), which reveal a compensation-like behaviour around $T_{comp}$= 50 K, where $M_s$ exhibits a dip, and $\mu_0 H_c$ shows a peak (see **Figure 4(b)**). Bulk LuIG garnet is not expected to possess a magnetic compensation point because $Lu^{3+}$ ions are nonmagnetic. However, in our films, the presence of a 3 nm-thick interfacial mixed $(Lu_{1-x}Gd_x)_3Fe_5O_{12}$ layer at the film/substrate interface contributes to the overall magnetic response, giving rise to a compensation point. Notice that the saturation magnetisation decreases across the compensation but does not go to zero, since we probe the whole layer magnetisation by VSM, i.e., both the Gd-doped LuIG region and the bulk LuIG region. The bulk LuIG layer remains magnetic over the entire temperature range, explaining why the total $M_s$ does not go to zero.

To further investigate the presence of compensation behaviour due to the Gd-doping, we performed interface-sensitive magnetoelectronic transport measurements. The sketch of the stack and the experimental configuration used to record the Hall voltage is shown in **Figure 4 (c)**. When a charge current is injected through Pt in the presence of a bias perpendicular magnetic field $\mu_0 H$, it gives rise to Hall resistivity $\rho_{xy}$. The total Hall resistivity is described as $\rho_{xy} = \rho_0 \mu_0 H + \rho_A M_z + \rho_{THE}$, where $\rho_0$ is the ordinary Hall coefficient, $\rho_A$ is the anomalous Hall coefficient, $M_z$ is the average $z$ component of the magnetisation unit vector, and $\rho_{THE}$ is the topological Hall coefficient. In HM/garnet heterostructures, the AHE signal is dominated by the spin-Hall anomalous Hall effect (SH-AHE) due to the occurrence of the spin-Hall effect via inverse SHE in the HM (Pt, Ta, W…) layer. In addition, a topological Hall (SH-THE) contribution can be detected in some cases [12], which arises from the interaction between non-collinear spin textures in the garnet film and the electrons from the heavy metal.

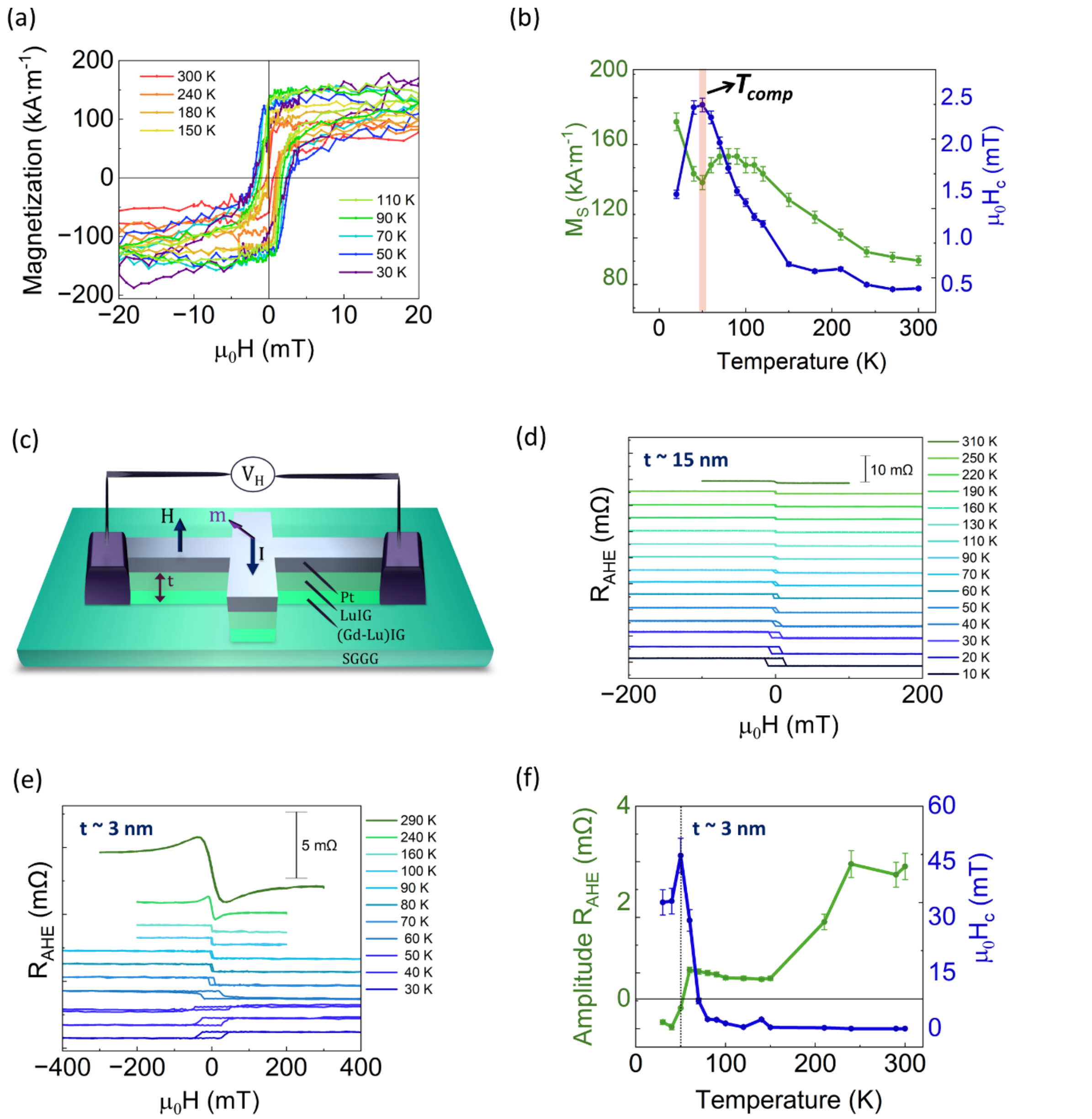

**Figure 4:** (a)Temperature-dependent OOP loops of the 20 nm thick layer with (b) saturation magnetization ($M_S$) and coercive field ($\mu_0 H_C$) as a function of temperature. (c) Schematic representation of the Pt/LuIG Hall bar device on a SGGG(111) substrate and magneto-transport measurements. Temperature dependence of the Hall resistance ($R_{AHE}$) as a function of out-of-plane applied magnetic field for several Pt (3 nm)/LuIG($t$) systems: (d) $t$ = 15 nm, (e) $t$ = 3 nm. (f) Amplitude of $R_{AHE}$ and the coercive field ($\mu_0 H_c$) as a function of temperature for $t$ = 3 nm, showing a compensation temperature ($T_M$) around $T$= 50 K. The linear contribution of the OHE in Pt has been subtracted from the data.

We measured the Hall resistance ($R_{AHE}$) of Pt (3 nm)/LuIG (t) heterostructures ($t$= 15 nm down to 3 nm) as a function of the OOP magnetic field for different temperatures.The anomalous Hall hysteresis loops, obtained after subtraction of the linear background arising from the ordinary Hall effect, are shown in **Figure 4 (d)** for the 15 nm thick LuIG layer. We observe typical sharp

AHE hysteresis loops as a function of the applied magnetic field, without any Hall sign change down to 10 K. This is explained by the fact that there is no contact between the Gd-rich LuIG layer and Pt and thus no Hall sign change related to magnetic compensation. If we decrease the LuIG film thickness, we expect that the Gd-rich LuIG layer close to the Pt interface can have an impact on the electronic transport. For an 8 nm LuIG film thickness, we still do not observe a change in the AHE sign, as shown in **Figure S4**. However, for an ultrathin film of 3 nm of LuIG, we observe a striking effect (see **Figure 4 (e and f)**). The Hall sign changes around $T$ = 50 K, along with a peak in the coercive field, exhibiting the compensation temperature ($T_{comp}$). At high temperatures ($T > T_{comp}$), the magnetization is strongly influenced by the tetrahedral $Fe^{3+}$ sublattice, while at low temperatures ($T < T_{comp}$), the magnetic moment of Gd ions becomes dominant, causing net magnetization to align with the direction of the Gd-containing sublattice. Thus, the interface-sensitive transport measurements provide direct proof of magnetic compensation arising from Gd diffusion in LuIG films on a few nanometers, which is consistent with chemical (EELS) and magnetic (VSM) analyses. More generally, the Gd diffusion (from a SGGG or GGG substrate) in garnet thin films due to a high growth temperature could explain discrepancies in previous reports, which showed a magnetic compensation in garnet films where no compensation temperature is usually expected in the bulk [5,12,35].

## 2.4 Observation of SH -THE-like effect in LuIG films

The topological Hall effect has often served as a signature for the detection of magnetic textures stabilized by Dzyaloshinskii–Moriya interactions in magnetic thin films for over a decade [36–40]. Several explanations have been proposed for the origin of DMI in centrosymmetric insulating magnetic garnets: interfacial orbital hybridization between the heavy metal and Fe ions of the garnet [37,41], orbital magnetism of rare-earth ions, garnet film thickness [42,43], or substrate-induced strain effect [36,44]. Previous studies on the interfacial DMI in garnet have predominantly focused on TmIG or YIG, and so far there has been no experimental observation of DMI in LuIG.

In order to investigate the possible presence of DMI which could be related to non-collinear structures in LuIG thin films, we have explored the AHE loops at higher temperatures. As detailed in **Figure 4(e)**, we can observe additional peaks in the AHE loops for different LuIG film thicknesses as a function of temperature. These features are often attributed to a topological Hall-like contribution and have been obtained after subtraction of the ordinary Hall and anomalous Hall effects. The anomalous Hall loop is therefore described by a hyperbolic tangent function given by: $\rho_{AHE}\ (\mu_0 H_Z) = \rho_{AHE\text{-}MAX}\ tanh\ [(H_Z - h_0)/h]$, where $\rho_{AHE}$ is the amplitude of SH-AHE, $\mu_0 H_z$ is the perpendicularly applied magnetic field, $\rho_{AHE\text{-}MAX}$ is the saturation value of AHE, $\mu_0 h_0$ is the coercive field, and $\mu_0 h$ is the width of switching. We fit the Hall data with the AHE function to obtain the SH-THE resistivity in various Pt/LuIG bilayers (see **Figure 5 (a)**).

Further, using the obtained SH-THE resistivity, we have plotted the SH-THE phase diagram as a function of temperature and applied perpendicular magnetic field (see **Figure 5 (b)** and **Figure S4)**. As the film thickness of LuIG increases, the SH-THE resistivity window shifts to higher temperatures. The data set for each thickness is fitted with a Gaussian function to determine the maximum SH-THE resistivity $\rho_{THE-max}$ for each sample, as shown in **Figure 5(c)**. We observe that the magnitude of $\rho_{THE-max}$ decreases with decreasing LuIG film thickness, contrary to previous reports that demonstrated an inverse dependence [36,39,45] with the REIG film thickness. Our results suggest that the observed behaviour cannot be explained by a pure interfacial mechanism. We speculate that biaxial strain and chemical inhomogeneity in our samples can break spatial symmetry, enabling the emergence of a net spin-orbit coupling that can interact with localised spins to stabilise DMI [46,47]. Therefore, different contributions at the HM/REIG and REIG/substrate interfaces can have an impact on the SH-THE-like effect and the magnitude of the DMI. However, a thorough understanding of the existence of non-collinear structures in LuIG/Pt bilayers (such as skyrmions) stabilized by DMI calls for further in-depth study combining magnetic imaging with rigorous magnetoelectronic transport response.

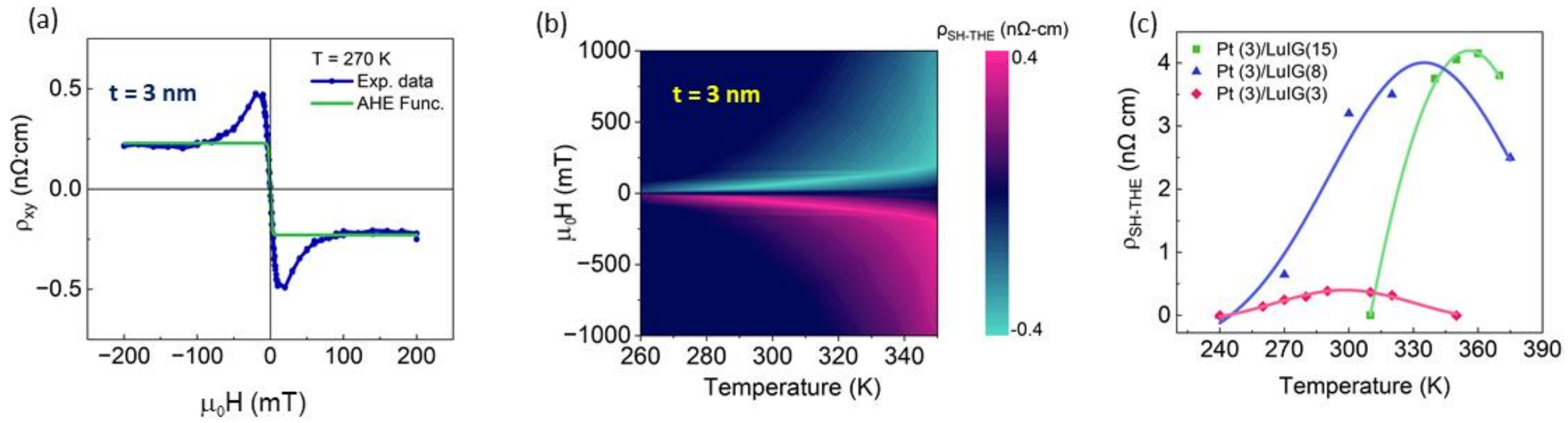

***Figure 5:*** Spin-Hall topological-like effect of Pt (3 nm)/LuIG(t)bilayers (a) Hall resistivity ($\rho_{xy}$) of a Pt (3 nm)/LuIG (3 nm) bilayer at $T$= 270 K. The green curve depicts the fit to the SH-AHE contribution modelled by the hyperbolic tangent function. (b) Magnetic field versus temperature phase diagram of the SH-THE resistivity ($\rho_{SH\text{-}THE}$) for a Pt (3 nm)/LuIG (3 nm) bilayer. (c) Evolution of the maximum value of $\rho_{SH\text{-}THE}$ with temperature for different film thicknesses of LuIG ($t$= 3 nm, 8 nm and 15 nm).

## 3. Conclusion

This work has presented a comprehensive study of epitaxial $Lu_3Fe_5O_{12}$(111) thin films grown on SGGG(111) substrates with a special attention given to the relationships between the chemical nature of the LuIG/SGGG interface and the physical properties. At first, we have reported an epitaxial strain-induced enhancement of perpendicular magnetic anisotropy in LuIG layers and LuIG/Pt bilayers with a low magnetic damping even for LuIG ultrathin films (down to 3 nm). Further, we have revealed Gd diffusion at the LuIG/SGGG interface during the high temperature growth by combining electron energy loss spectroscopy, magnetic and interface-sensitive electronic transport (Anomalous Hall effect). The detection of a 3 nm thick Gd-rich region in LuIG films results in: *(i)* the presence of a compensation temperature around 50 K as observed in magnetometry and Hall measurements and *(ii)* the composition gradient-induced structural inversion symmetry breaking which could modulate the DMI and contribute to SH-THE-like effects observed in the high temperature regime. The presence of a compensation temperature, along with a low damping and possible non colinear spin textures in LuIG garnets thin films paves the way for advancements in tunable low-power spintronic devices.

## 4. Experimental Section

## 4.1. Sample Preparation

A polycrystalline $Lu_3Fe_5O_{12}$ (LuIG) target is prepared by solid-state reaction from a stoichiometric mixture of $Lu_2O_3$ (99.9% purity) and $Fe_2O_3$ (99.9% purity) powders sintered at 1175°C, 1350°C, and 1400°C for 7 h, 7 h, and 5 h, respectively, for pulsed laser deposition (PLD) of LuIG film. The target is ablated using a nanosecond laser beam (Nd: YAG laser) of 355 nm wavelength with fluences varying between 1.5 and 0.90 J $cm^{-2}$ and a repetition rate of 2.5 Hz during the film growth. Epitaxial LuIG films of different thicknesses are grown on 5 mm x 5 mm (111)-oriented single-crystalline $Gd_3(Mg,Zr)Ga_5O_{12}$ (SGGG) substrates. The samples are cleaned using acetone, isopropanol, and deionised water before being transferred into the PLD chamber. The base pressure in the PLD chamber is less than $10^{-8}$ mbar. The substrates are heated to 700°C, and the growth is carried out at an oxygen partial pressure of 2.5 $10^{-1}$ mbar. A high-pressure reflection high-energy electron diffraction (RHEED) gun is used in real time to monitor on a florescent screen the crystalline structure of the atomic planes along the [1-10] and [11-2] azimuthal directions. A 3 nm-thick Pt layer is deposited *in-situ* on LuIG films at room temperature using an e-beam gun with an evaporation rate around xx measured by a quartz balance.

## 4.2. Structural Characterisation

The film thickness is checked ex-situ by X-ray reflectivity (XRR). The Kiessig oscillations, are adjusted by using the SimulReflec software with density, atomic diffusion factors, roughness and film thickness as fitting parameters.
Atomic force microscopy (AFM) topographic images are obtained using the Asylum Research MFP-3D microscope in tapping mode to investigate the surface morphology of the substrate and the groughness of the LuIG films.
The crystalline phase of the LuIG thin films is probed by high-resolution X-ray diffraction (XRD) using a 5-circle diffractometer (Rigaku SmartLab) running in the parallel beam geometry with $\lambda = 1.5406$ Å (Cu K$\alpha$).

## 4.3. Transmission Electron Microscopy

The cross-section specimen was prepared using Focused Ion Beam (FIB, Ga + source, Helios 600i). High-resolution transmission electron microscopy (HRTEM) images are recorded using FEI Tecnai G2 microscope with CEOS image corrector operating at 200 kV. The EELS acquisition has been done on the same microscope in STEM mode using the GIF TRIDIEM EELS spectrometer..

## 4.4. Static and dynamic magnetic Characterisation

The *M-H* hysteresis loops are obtained using a Quantum Design Physical Properties Measurement System with the VSM option (7 T PPMS-VSM, sensitivity: 10-9 A.$m^2$). The *M-H* curves can be obtained between 380 K and 20 K. The samples are placed in two different geometries in order to obtain *M-H* curves with the magnetic field applied parallel or perpendicular to the film surface.
The magnetic anisotropy and dynamic properties of the grown film are investigated by a broadband FMR setup using a microwave stripline of 0.5 mm width at room temperature. To sensitively probe the absorbed power at FMR, the swept in-plane bias magnetic field is modulated at 35 Hz (modulation amplitude equal to 0.1 - 0.2 mT), and the diode voltage is lock-in detected.

### 4.5. Device Preparation and Electrical Measurements

A Hall bar geometry of $150 \times 50\ \mu m^2$ was patterned using standard photolithography techniques and a dry Ar ion etching process on Pt/LuIG samples. The Hall measurements are carried out in the Quantum Design Physical Property Measurement System (PPMS) using a DC current of 300 μA in a field range of +/- 1 T at various temperatures (10 K-370 K).

## Acknowledgements

The authors thank André Thiaville and Nicolas Reyren for useful conversations. This work was financially supported by the ANR Demiurge project, No. ANR-22-CE30-0014. The authors also acknowledge the support of the France 2030 government investment plan managed by the French National Research Agency under grant reference PEPR SPIN – SPINMAT ANR-22-EXSP-0007. The authors also acknowledge the support from UPSTAIRS Grant number: ANR-10-LABX-0039-PALM.

## Conflicts of Interest

The authors declare no conflicts of interest

## Data Availability Statement

The data needed to evaluate the conclusions of this study are available from the corresponding author upon reasonable request.

## References:

[1] C. Hauser, T. Richter, N. Homonnay, C. Eisenschmidt, M. Qaid, H. Deniz, D. Hesse, M. Sawicki, S. G. Ebbinghaus, G. Schmidt, *Sci Rep* **2016**, *6*, 20827.

[2] A. Quindeau, C. O. Avci, W. Liu, C. Sun, M. Mann, A. S. Tang, M. C. Onbasli, D. Bono, P. M. Voyles, Y. Xu, J. Robinson, G. S. D. Beach, C. A. Ross, *Advanced Electronic Materials* **2017**, *3*, 1600376.

[3] K. Ganzhorn, S. Klingler, T. Wimmer, S. Geprägs, R. Gross, H. Huebl, S. T. B. Goennenwein, *Appl. Phys. Lett.* **2016**, *109*, 022405.

[4] R. Kumar, S. N. Sarangi, D. Samal, Z. Hossain, *Phys. Rev. B* **2021**, *103*, 064421.

[5] S. Ding, Z. Liang, C. Yun, R. Wu, M. Xue, Z. Lin, A. Ross, S. Becker, W. Yang, X. Ma, D. Chen, K. Sun, G. Jakob, M. Kläui, J. Yang, *Phys. Rev. B* **2021**, *104*, 224410.

[6] S. Pandey, A. Barbier, A. Forget, B. Sarpi, F. Maccherozzi, R. Sant, N. B. Brookes, J.-B. Moussy, P. V. B. Pinho, *ACS Appl. Electron. Mater.* **2026**, *8*, 2391.

[7] J. J. Bauer, P. Quarterman, A. J. Grutter, B. Khurana, S. Kundu, K. A. Mkhoyan, J. A. Borchers, C. A. Ross, *Phys. Rev. B* **2021**, *104*, 094403.

[8] M. B. Jungfleisch, V. Lauer, R. Neb, A. V. Chumak, B. Hillebrands, *Appl. Phys. Lett.* **2013**, *103*, 022411.

[9] K. Uchida, T. Nonaka, T. Kikkawa, Y. Kajiwara, E. Saitoh, *Phys. Rev. B* **2013**, *87*, 104412.

[10] Y. Li, Y. Duan, M. Wang, L. Lang, Y. Zhang, K. Shen, Z. Shi, S.-M. Zhou, *npj Spintronics* **2025**, *3*, 15.

[11] J. Kim, P. C. Van, J.-R. Jeong, H. Jin, *Advanced Functional Materials* **2026**, *36*, e27195.

[12] Q. Shao, A. Grutter, Y. Liu, G. Yu, C.-Y. Yang, D. A. Gilbert, E. Arenholz, P. Shafer, X. Che, C. Tang, M. Aldosary, A. Navabi, Q. L. He, B. J. Kirby, J. Shi, K. L. Wang, *Phys. Rev. B* **2019**, *99*, 104401.

[13] C. O. Avci, E. Rosenberg, M. Baumgartner, L. Beran, A. Quindeau, P. Gambardella, C. A. Ross, G. S. D. Beach, *Appl. Phys. Lett.* **2017**, *111*, 072406.

[14] S. Vélez, J. Schaab, M. S. Wörnle, M. Müller, E. Gradauskaite, P. Welter, C. Gutgsell, C. Nistor, C. L. Degen, M. Trassin, M. Fiebig, P. Gambardella, *Nat Commun* **2019**, *10*, 4750.

[15] Z. Ye, Z. Chen, Z. Chen, W. Jia, T. Gao, L. Liu, H. Zheng, Q. Zeng, Q. Wang, N. Wang, B. Xiang, T. Lin, M. Qiu, S. Li, J. Shi, Z. Hou, K. Ando, H. An, *Advanced Electronic Materials* **2022**, *8*, 2100590.

[16] S. Husain, N. F. Prestes, O. Fayet, S. Collin, F. Godel, E. Jacquet, T. Denneulin, R. E. Dunin-Borkowski, A. Thiaville, M. Bibes, H. Jaffrès, N. Reyren, A. Fert, J.-M. George, *Nano Lett.* **2024**, *24*, 2743.

[17] L. Caretta, E. Rosenberg, F. Büttner, T. Fakhrul, P. Gargiani, M. Valvidares, Z. Chen, P. Reddy, D. A. Muller, C. A. Ross, G. S. D. Beach, *Nat Commun* **2020**, *11*, 1090.

[18] E. I. Ahamed, M. S. Sarker, H. Yamahara, S. Tang, K. Morikawa, M. Seki, H. Tabata, *APL Mater.* **2025**, *13*, 041114.

[19] H. Yamahara, M. Seki, H. Tabata, *Journal of Magnetism and Magnetic Materials* **2020**, *501*, 166437.

[20] J. Bensmann, R. Schmidt, K. O. Nikolaev, D. Raskhodchikov, S. Choudhary, R. Bhardwaj, S. Taheriniya, A. Varri, S. Niehues, A. El Kadri, J. Kern, W. H. P. Pernice, S. O. Demokritov, V. E. Demidov, S. Michaelis de Vasconcellos, R. Bratschitsch, *Nat. Mater.* **2025**, *24*, 1920.

[21] M. J. Roos, P. Quarterman, J. Ding, M. Wu, B. J. Kirby, B. L. Zink, *Phys. Rev. Mater.* **2022**, *6*, 034401.

[22] A. Mitra, O. Cespedes, Q. Ramasse, M. Ali, S. Marmion, M. Ward, R. M. D. Brydson, C. J. Kinane, J. F. K. Cooper, S. Langridge, B. J. Hickey, *Sci Rep* **2017**, *7*, 11774.

[23] J. F. K. Cooper, C. J. Kinane, S. Langridge, M. Ali, B. J. Hickey, T. Niizeki, K. Uchida, E. Saitoh, H. Ambaye, A. Glavic, *Phys. Rev. B* **2017**, *96*, 104404.

[24] C. Dubs, O. Surzhenko, R. Thomas, J. Osten, T. Schneider, K. Lenz, J. Grenzer, R. Hübner, E. Wendler, *Phys. Rev. Mater.* **2020**, *4*, 024416.

[25] T. Yamagishi, J. Awaka, Y. Kawashima, M. Uemura, S. Ebisu, S. Chikazawa, S. Nagata *, *Philosophical Magazine* **2005**, *85*, 1819.

[26] E. I. Ahamed, H. Yamahara, M. S. Sarker, H. Li, K. Morikawa, K. Yamagami, M. Kobayashi, M. Seki, H. Tabata, *Advanced Electronic Materials* **2025**, *11*, 2400735.

[27] C. Holzmann, A. Ullrich, O.-T. Ciubotariu, M. Albrecht, *ACS Appl. Nano Mater.* **2022**, *5*, 1023.

[28] J. M. Gomez-Perez, S. Vélez, L. McKenzie-Sell, M. Amado, J. Herrero-Martín, J. López-López, S. Blanco-Canosa, L. E. Hueso, A. Chuvilin, J. W. A. Robinson, F. Casanova, *Phys. Rev. Appl.* **2018**, *10*, 044046.

[29] Magnetic Properties and Growth-Induced Anisotropy in Yttrium Thulium Iron Garnet Thin Films - Rosenberg - 2021 - Advanced Electronic Materials - Wiley Online Library, .

[30] J. Ding, C. Liu, Y. Zhang, U. Erugu, Z. Quan, R. Yu, E. McCollum, S. Mo, S. Yang, H. Ding, X. Xu, J. Tang, X. Yang, M. Wu, *Phys. Rev. Appl.* **2020**, *14*, 014017.

[31] I. N. Bhatti, I. N. Bhatti, C. Carrétéro, V. Cros, P. Bortolotti, R. Lebrun, S. Mantion, N. Reyren, A. Anane, *Phys. Rev. Mater.* **2026**, *10*, 064410.

[32] Low-damping sub-10-nm thin films of lutetium iron garnet grown by molecular-beam epitaxy | Applied Physics Letters | AIP Publishing, .

[33] B. Heinrich, C. Burrowes, E. Montoya, B. Kardasz, E. Girt, Y.-Y. Song, Y. Sun, M. Wu, *Phys. Rev. Lett.* **2011**, *107*, 066604.

[34] A. Hamadeh, O. d'Allivy Kelly, C. Hahn, H. Meley, R. Bernard, A. H. Molpeceres, V. V. Naletov, M. Viret, A. Anane, V. Cros, S. O. Demokritov, J. L. Prieto, M. Muñoz, G. de Loubens, O. Klein, *Phys. Rev. Lett.* **2014**, *113*, 197203.

[35] C. C. Soares, T. J. A. Mori, F. Béron, J. S. Moodera, J. C. Cezar, J. Brandão, G. Vilela, *ACS Appl. Nano Mater.* **2025**, *8*, 14567.

[36] A. S. Ahmed, A. J. Lee, N. Bagués, B. A. McCullian, A. M. A. Thabt, A. Perrine, P.-K. Wu, J. R. Rowland, M. Randeria, P. C. Hammel, D. W. McComb, F. Yang, *Nano Lett.* **2019**, *19*, 5683.

[37] A. J. Lee, A. S. Ahmed, J. Flores, S. Guo, B. Wang, N. Bagués, D. W. McComb, F. Yang, *Phys. Rev. Lett.* **2020**, *124*, 107201.

[38] Z. Xu, Y. Zhu, Y. Wang, X. Li, Q. Liu, K. Chen, J. Wang, Y. Jiang, L. Chen, *Advanced Science* **2024**, *11*, 2403852.

[39] Q. Shao, Y. Liu, G. Yu, S. K. Kim, X. Che, C. Tang, Q. L. He, Y. Tserkovnyak, J. Shi, K. L. Wang, *Nat Electron* **2019**, *2*, 182.

[40] Investigation of the Role of Rare-Earth Elements in Spin-Hall Topological Hall Effect in Pt/Ferrimagnetic-Garnet Bilayers | Nano Letters | ACS Publications, .

[41] J. Huang, Y. Li, S. Hu, J. Ke, J. W. Cai, Z. Zhang, Y. Liu, *Phys. Rev. B* **2026**, *113*, 144437.

[42] S. Ding, A. Ross, R. Lebrun, S. Becker, K. Lee, I. Boventer, S. Das, Y. Kurokawa, S. Gupta, J. Yang, G. Jakob, M. Kläui, *Phys. Rev. B* **2019**, *100*, 100406.

[43] H. Wang, J. Chen, T. Liu, J. Zhang, K. Baumgaertl, C. Guo, Y. Li, C. Liu, P. Che, S. Tu, S. Liu, P. Gao, X. Han, D. Yu, M. Wu, D. Grundler, H. Yu, *Phys. Rev. Lett.* **2020**, *124*, 027203.

[44] Z. Xu, Q. Liu, Y. Ji, X. Li, J. Li, J. Wang, L. Chen, *ACS Appl. Mater. Interfaces* **2022**, *14*, 16791.

[45] M. Loyal, A. Akashdeep, E. Mangini, E. Galíndez-Ruales, M. Eich, N. Wang, Q. Lan, L. Jin, R. Dunin-Borkowski, T. Kuschel, M. Kläui, G. Jakob, *Phys. Rev. Mater.* **2026**, *10*, 054417.

[46] J. Liang, M. Chshiev, A. Fert, H. Yang, *Nano Lett.* **2022**, *22*, 10128.

[47] S. Krishnia, E. Haltz, L. Berges, L. Aballe, M. Foerster, L. Bocher, R. Weil, A. Thiaville, J. Sampaio, A. Mougin, *Phys. Rev. Appl.* **2021**, *16*, 024040.

**ToC figure**

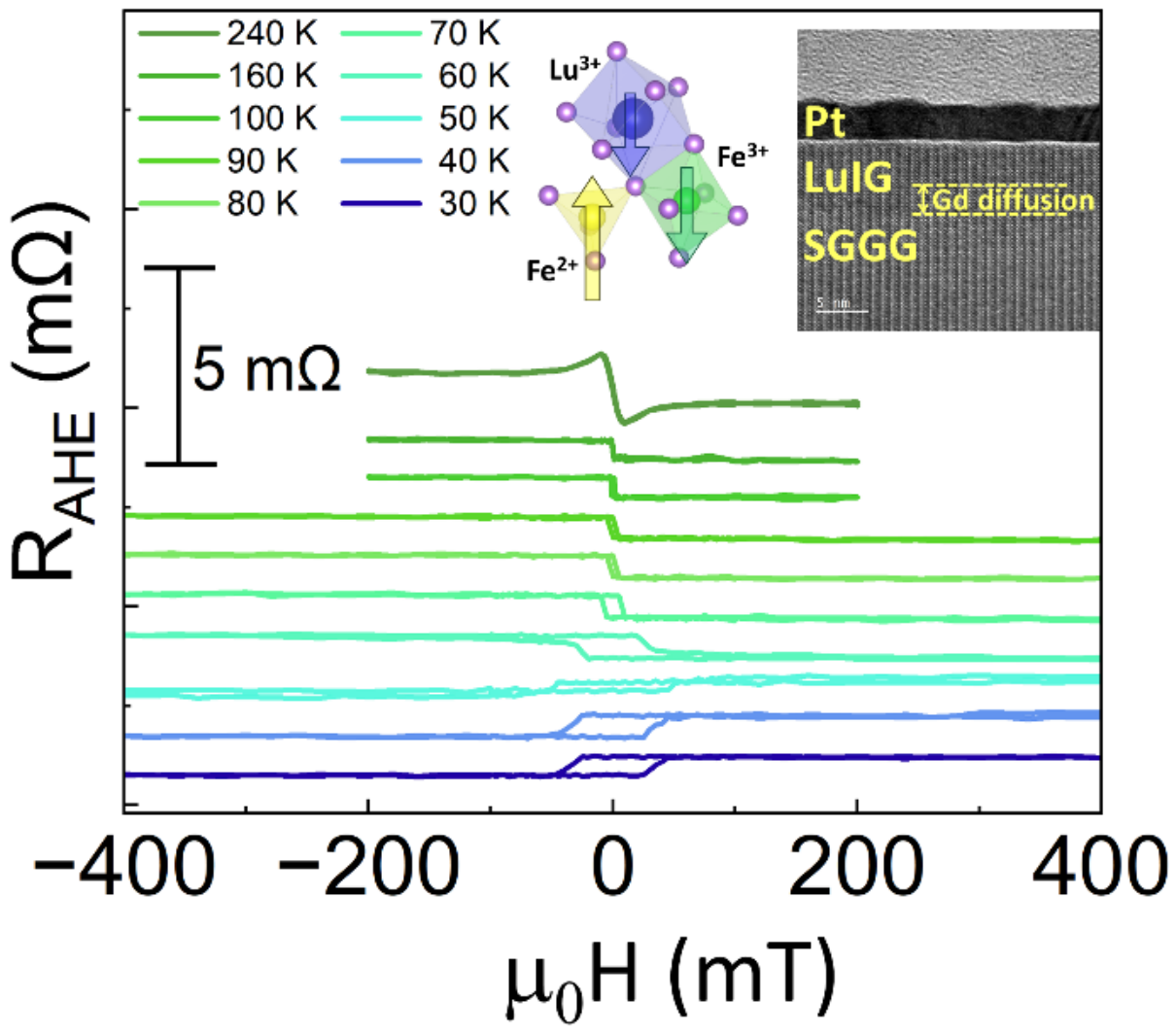

240 K
160 K
100 K
90 K
80 K
70 K
60 K
50 K
40 K
30 K
Lu3+
Fe3+
Fe2+
Pt
LuIG
Gd diffusion
SGGG
5 mΩ
R_AHE (mΩ)
−400
−200
0
200
400
μ0H (mT)

# Supporting Information

**Interfacial origin of magnetic compensation and topological-like Hall effect in crystalline lutetium iron garnet thin films**

*Satakshi Pandey*[1], *Anne Forget*[1], *Sarah Clement*[2], *Christophe Gatel*[2], *Bénédicte Warot-Fonrose*[2], *Grégoire de Loubens*[1], *Jean-Baptiste Moussy*[1]

[1] Université Paris-Saclay, CEA/CNRS, Service de Physique de l'Etat Condensé, 91191 Gif-sur-Yvette, France.

[2] Université de Toulouse, CNRS, CEMES, 31055 Toulouse, France.

## 1. Structural characterisation of LuIG films

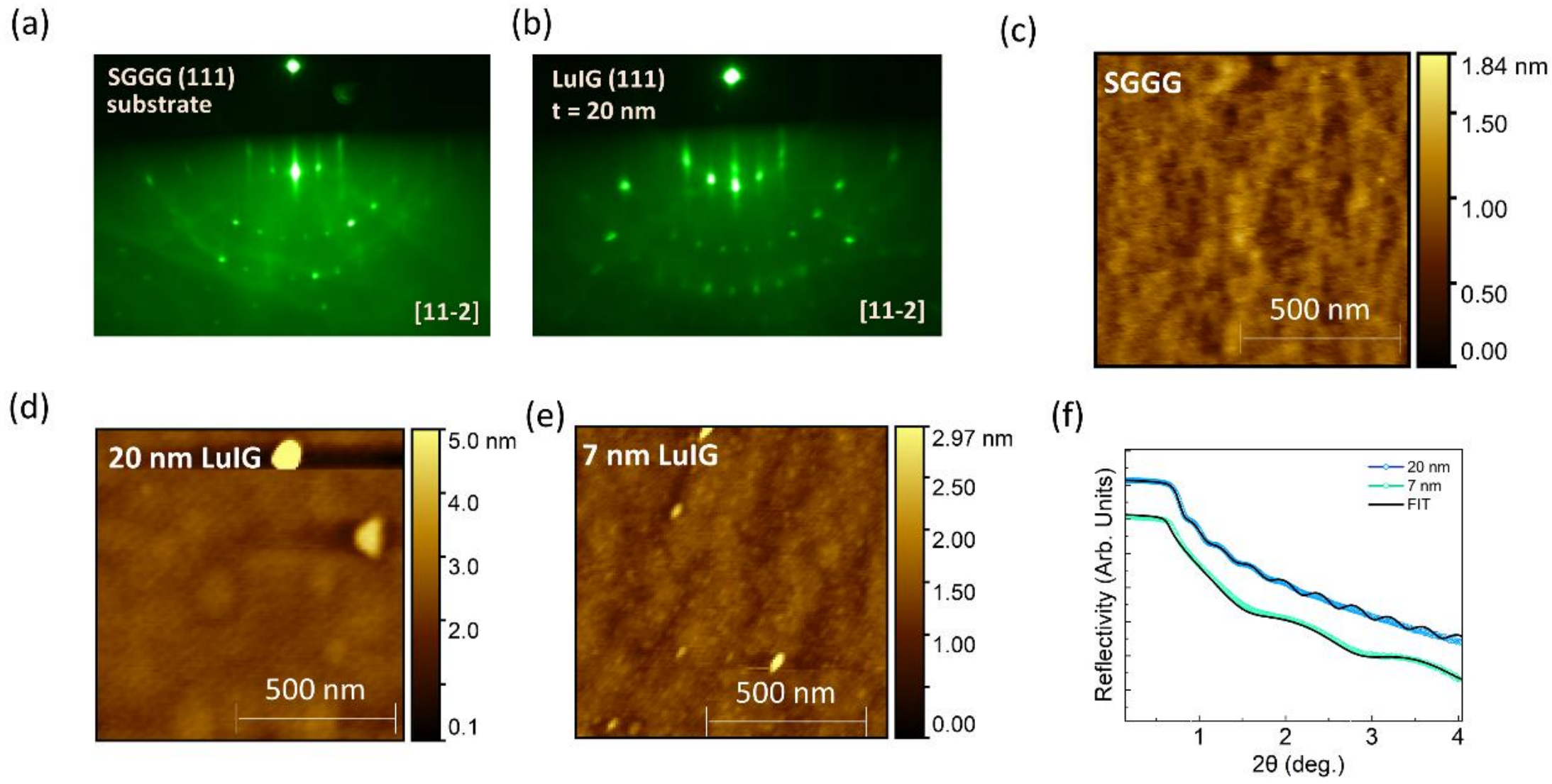

**Figure S1**: RHEED diffraction patterns along the $[11\bar{2}]$ azimuth of the SGGG (111) substrate and the deposited (a) 20 nm and (b) 7 nm LuIG(111) thin film. AFM topographic images (1 µm × 1 µm) acquired in tapping mode of the (c) SGGG substrate and the deposited (d) 20 nm and (e) 7 nm LuIG thin films. (f) XRR curves of 20 nm (blue) and 7 nm (green) LuIG films with fittings using Parratt's formalism.

## 2. FMR characterisation of LuIG films

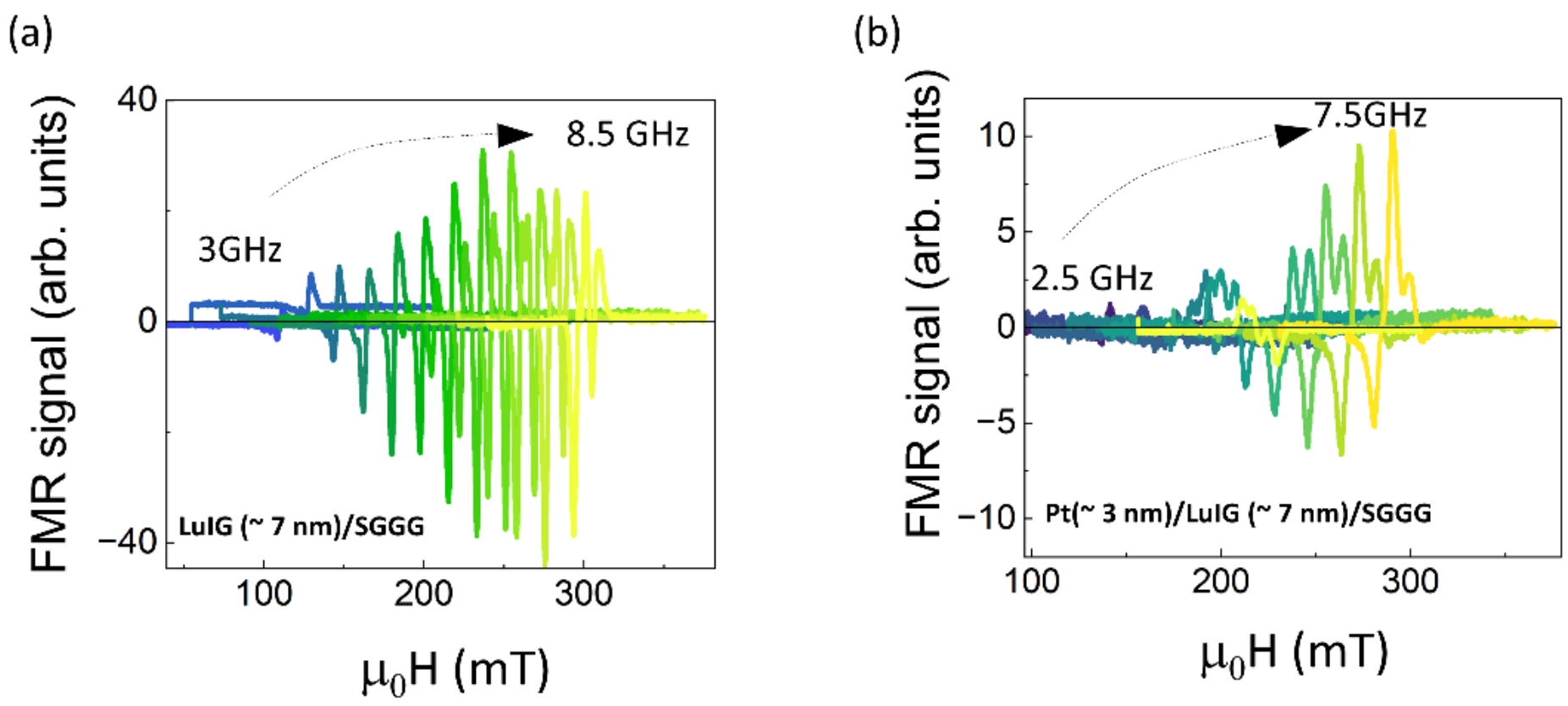

**Figure S2:** Room-temperature Ferromagnetic resonance (FMR) spectra of LuIG/SGGG samples with thicknesses of (a) 7 nm and (b) 7 nm, capped with Pt, measured at different frequencies.

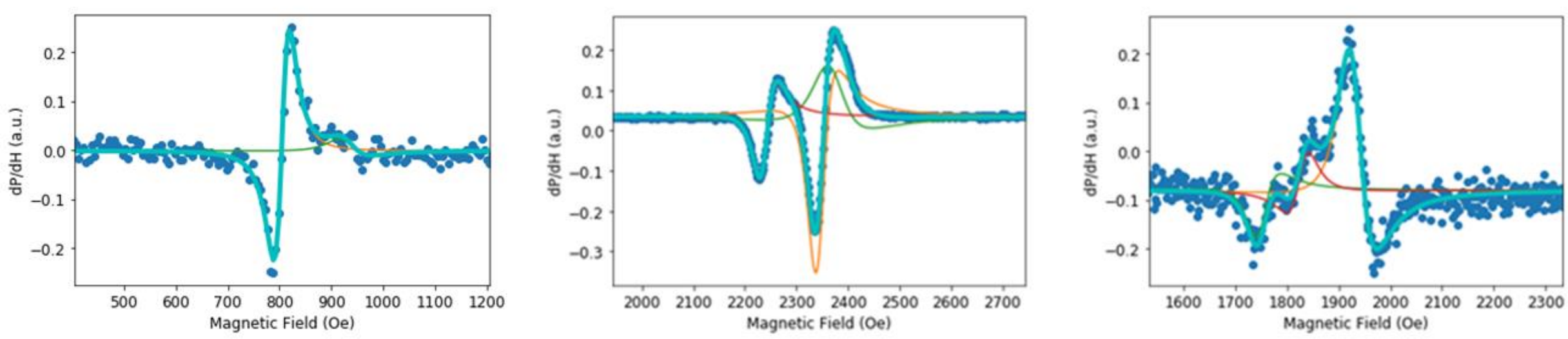

**Figure S3:** Examples of multiple Lorentzian fits of FMR spectra. (a) LuIG(20nm) at 2.5 GHz; (b) LuIG(7nm) at 6.5 GHz; (c) Pt(3nm)/LuIG(7nm) at 4.5 GHz. Dots correspond to experimental data, the thin orange, green and red lines to the fitted peaks, and the thick blue line to the overall fit.

### 3. AHE and SH-THE-like effects in LuIG films

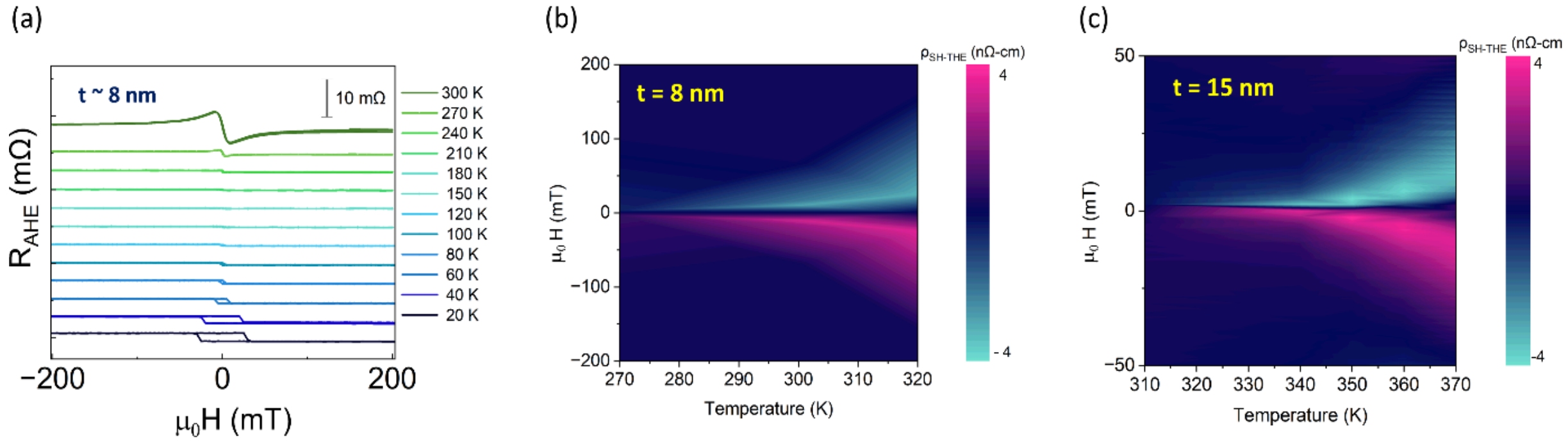

**Figure S4:** (a) Temperature dependence of the Hall resistance ($R_{AHE}$) as a function of the out-of-plane applied magnetic field for Pt (3 nm)/LuIG(8). Topological spin-texture phase diagrams depicting the magnetic field versus temperature ($T$) phase diagram of the SH-THE resistivity for (b) Pt (3 nm)/ LuIG (8 nm), (c) Pt (3 nm)/ LuIG (15 nm)